\documentclass{article}
\usepackage{spconf,amsmath,graphicx}
\usepackage{hyperref}

\usepackage[T1]{fontenc}
\usepackage[utf8]{inputenc}
\usepackage{amsmath} 
\usepackage{amsfonts} 
\usepackage{amssymb} 
\usepackage{amsthm}
\usepackage{color}
\usepackage{url}
\usepackage{subcaption}
\usepackage{float}
\usepackage[linesnumbered,ruled,vlined]{algorithm2e}
\usepackage{verbatim}
\usepackage{multirow}
\usepackage{graphicx}
\usepackage{comment}
\SetKwInput{KwInput}{Input}                
\SetKwInput{KwOutput}{Output}              
\SetKwInput{Kwinitialize}{Initialization}              

\usepackage{cite}
\renewcommand{\vec}[1]{{\bf{#1}}} 
\newcommand{\vecgreek}[1]{{\boldsymbol{#1}}} 
\newcommand{\tran}{^{\mbox{\scriptsize T}}}
\newcommand{\herm}{^{\mbox{\scriptsize H}}}

\newcommand{\fro}[1]{\| #1\|_{\mathrm{F}}^2}
\newcommand{\norm}[1]{\| #1\|_2}

\title{Iterative Reweighted Algorithms for Joint User Identification and Channel Estimation in Spatially Correlated Massive MTC}
\name{Hamza Djelouat, Markus Leinonen, and Markku Juntti \thanks{This work has been financially supported in part by the Academy of Finland (ROHM project, grant 319485) and Academy of Finland 6Genesis Flagship (grant 318927). The work of M. Leinonen has also been financially supported in part by Infotech Oulu and the Academy of Finland (grant 323698).}}
\address{Centre for Wireless Communications -- Radio Technologies, FI-90014, University of Oulu, Finland}

\begin{document}
\ninept
\maketitle
\begin{abstract}
Joint user identification and channel estimation (JUICE) is a main challenge in grant-free massive machine-type communications (mMTC). The sparse pattern in users' activity allows to solve the JUICE as a compressed sensing problem in a multiple measurement vector (MMV) setup. This paper addresses the JUICE under the practical spatially correlated fading channel. We formulate the JUICE as an iterative reweighted $\ell_{2,1}$-norm optimization. We develop a computationally efficient alternating direction method of multipliers (ADMM) approach to solve it. In particular,  by leveraging the second-order statistics of the channels, we reformulate the JUICE problem to exploit the covariance information and we derive its ADMM-based solution. The simulation results highlight the significant improvements brought by the proposed approach in terms of channel estimation and activity detection performances.
\end{abstract}
\section{Introduction}
\label{sec:intro}
The demand for internet of things (IoT) applications drives the deployment of massive machine-type communications  (mMTC)  as a major use case in 5G wireless technologies. mMTC implies sporadic uplink  communication from a massive number of IoT devices, called user equipments (UEs). Therefore,  communications with low signalling overhead is needed. To this end, grant-free access has been identified as a key enabler for mMTC \cite{cirik2019toward}.  It requires joint identification of the active UEs and estimation of their channel state information (CSI), known as the \textit{joint user identification and channel estimation} (JUICE) problem.

The sparse user activity pattern along with the multi-antenna base station (BS) setup motivates the formulation of JUICE as a compressed sensing (CS) \cite{Candes-Romberg-Tao-06} problem in a multiple measurement vector (MMV) setup.
The optimal solution for sparse signal recovery requires solving an NP-hard $\ell_0$-norm minimization problem. Therefore,  several approaches have been proposed to overcome this limitation, e.g., mixed norm minimization \cite{steffens2018compact} (and the references therein), iterative algorithms \cite{tropp2006algorithms,donoho2009message}, and sparse Bayesian learning (SBL) \cite{wipf2007empirical}. 

In the context of grant-free based JUICE, the existing works  focus on techniques based on greedy algorithms \cite{jeong2018map,du2018block},  approximate message passing  \cite{chen2018sparse,liu2018massive,senel2018grant,ke2020compressive}, SBL \cite{zhang2017novel}, and  maximum likelihood estimation \cite{chen2019covariance}.  Nevertheless, since the aforementioned works assume that the channel components are independent, the performance of JUICE may deteriorate as this assumption is not always practical \cite{bjornson2016massive}.


In this paper, we formulate the JUICE as an iterative reweighted $\ell_{2,1}$-norm minimization. While the $\ell_1$-norm penalty follows from the conventional approximation of $\ell_0$-norm to relax the JUICE into a tractable convex problem, 
the reweighting compensates for the main
difference between the $\ell_1$- and $\ell_0$-norms: the dependency on coefficients' amplitude \cite{candes2008enhancing}. 
Moreover, differently from considering uncorrelated channels as in \cite{jeong2018map,du2018block,chen2018sparse,liu2018massive,senel2018grant,zhang2017novel,chen2019covariance}, we address the JUICE in the more practical spatially correlated   multiple-input multiple-output (MIMO) channels. In such models, the channel spatial correlation varies slowly compared to the channel realizations, hence, the channel can be estimated with high accuracy in practice \cite{Li-etal15}. Subsequently, the  spatial correlation information can be exploited to enhance the JUICE performance.

The main contributions of this paper are summarized as follows. First,  when the second-order statistics of the channels  are not available, we formulate the JUICE as an iterative reweighted $\ell_{2,1}$-norm minimization and we derive a computationally efficient solution based on alternating direction method of multipliers (ADMM) \cite{boyd2011distributed} by providing a closed-form expression to each sub-problem at each iteration. Second, when the BS knows the channels' second-order statistics, we augment the optimization problem with a penalty term on the deviation of the sample covariance matrices of the estimated channels from their respective true covariance matrices. Furthermore, once the active UEs are  identified, a minimum mean square error (MMSE) estimator is deployed to improve channel estimation. The proposed approaches are empirically shown to significantly improve the JUICE  performance.
%
\vspace{-.2cm}
\section{System Model}
\label{sec::1}
Consider a single-cell uplink network consisting of a  set of $N$ single-antenna UEs, $\mathcal{N}=\{1,\ldots\!,N\}$, communicating with  a  BS  equipped with a uniform linear array (ULA) containing $M$ antennas. We consider a block fading channel  over each coherence period $T_{\mathrm{c}}$. 
The channel response ${\vec{h}_{i} \in\mathbb{C}^{M}}\!$  between the $i$th UE and the BS is given as
\begin{equation}\footnotesize
    \vec{h}_i=\frac{1}{\sqrt{P_i}}\sum_{p=1}^{P_i}\omega_{i,p}\vec{a}(\psi_{i,p}), \;\;\;\forall i \in \mathcal{N},
    \label{eq::ch}
\end{equation}
where $P_i$ is the number of physical signal paths, $\omega_{i,p} \in \mathbb{C}$ accounts for the  $p$th path gain and $\vec{a}(\psi_{i,p}) \!\in \mathbb{C}^{M}$ is the array response of the ULA given as $[\vec{a}(\psi_{i,p})]_m\!= \!e^{-j (m-1) 2\pi\Delta_\mathrm{r}\cos(\psi_{i,p})}$, \!$m=1,\ldots,M$, where $\Delta_\mathrm{r}$ is the normalized space between each pair of BS antennas, and $\psi_{i,p} $ is the angle of arrival of the $p$th path~\cite{bjornson2019massive}. We focus on the case of a limited angular spread (i.e., highly directive channel).

At each coherence interval $T_\mathrm{c}$, a new and  independent  channel realization $\vec{h}_i$ in \eqref{eq::ch} is observed. The channels are considered to be wide-sense stationary \cite{Li-etal15}, i.e., the channel covariance matrix of the $i$th UE, denoted as  $\vec{R}_i=\mathbb{E}[\vec{h}_i\vec{h}_i\herm] \in \mathbb{C}^{M\times M}$, varies in a slower time-scale  compared to the channel realizations and it remains fixed for $\tau_\mathrm{s}$ coherence intervals, where $\tau_\mathrm{s}$  can be on the order of thousands \cite{bjornson2016massive,sanguinetti2019towards}. We assume the  common convention  that the covariance  matrices  $\{\vec{R}_i\}_{i=1}^N$  are known by the BS \cite{Li-etal15}. 


Due to the sporadic nature of mMTC, only ${K\ll N}$ UEs are active at each  $T_{\mathrm{c}}$.   Therefore, for coherent data detection, the active UEs have to be detected and their channels have to be estimated. To this end, the BS assigns to each UE $i\in \mathcal{N}$  a unit-norm pilot sequence $\vecgreek{\phi}_i \in \mathbb{C}^{\tau_{\mathrm{p}}}$. To mitigate the channel gain difference between the UEs, a power control policy is deployed such that UE $i$  transmits with a power $\rho_i$ that is inversely proportional to the average channel gain \cite{bjornson2016massive}. We define the pilot matrix as $\vec{\Phi}=[\vecgreek{\phi}_1,\ldots,\vecgreek{\phi}_N]\in \mathbb{C}^{\tau_\mathrm{p}\times N}$ and  the effective channel  matrix as $\vec{X}=[\vec{x}_1,\ldots,\vec{x}_{N}] \in \mathbb{C}^{M\times N}$, where $\vec{x}_i={\gamma_i \sqrt{\rho_i}\vec{h}_i}$ is the effective channel for $i$th UE  and $\gamma_i$ is an activity indicator, defined as
\begin{equation*}\footnotesize
\gamma_i =
\begin{cases}
1, & i \in \mathcal{S}\\ 
0 , & \text{otherwise},
\end{cases}
\end{equation*}
where $\mathcal{S}\subseteq \mathcal{N}$,  ${|\mathcal{S}|=K}$, denotes the set of active UEs.


During each $T_{\mathrm{c}}$, the $K$   active UEs  transmit their pilot sequences to the BS, and the  received pilot signal $\vec{Y} \in \mathbb{C}^{\tau_{\mathrm{p}}\times M}$ is given by
 \begin{equation}\footnotesize
     \vec{Y}= \vec{\Phi} \vec{X}\tran+ \vec{W},
     \label{eq::Y_matrix}
\end{equation}
where   $\vec{W} \in \mathbb{C}^{\tau_{\mathrm{p}}\times M}$ is additive white Gaussian noise with independent and identically distributed (i.i.d.) elements as $\mathcal{CN}(0,\,\sigma^{2})$,


\vspace{-.1cm}

\section{Proposed Solution via iterative approach}

\vspace{-.1cm}
\subsection{JUICE via Reweighted $\ell_{2,1}$-Norm Minimization}
\label{IRW-ADMM}
Since the columns of the effective channel matrix $\vec{X} $  corresponding to the inactive UEs are zero, $\vec{X}\tran$ has a \emph{row-sparse} structure. Thus,  JUICE can be modeled as a joint sparse  MMV recovery problem. The canonical form of optimal sparse recovery  requires solving a combinatorial NP-hard   $\ell_0$-norm minimization problem. Thus, a convex relaxation in the form of $\ell_{2,1}$-norm is considered in practice to obtain a computationally tractable problem, formulated as
\begin{equation}\footnotesize
\min _{\vec{X} }\frac{1}{2} \|\vec{\Phi} \vec{X}\tran - \vec{Y}\|_{\mathrm{F}}^2 + \beta_1 \|\vec{X}\tran\|_{2,1}.\label{eq::l_1}
\end{equation}
However, unlike the democratic $\ell_0$-norm  where the  non-zero coefficients are penalized equally, $\ell_1$-norm is biased toward larger magnitudes, i.e.,  coefficients with large amplitude are penalized more heavily than smaller ones \cite{candes2008enhancing}. 
Therefore, striving for a better recovery, we use the \emph{log-sum} penalty  to relax the  $\ell_{0}$-norm  as 
\begin{equation}\vspace{-.05cm}
\footnotesize\begin{array}{ll}
     & \displaystyle\min_{\vec{X},\vec{u} }\displaystyle\frac{1}{2}\|\vec{\Phi} \vec{X}\tran - \vec{Y}\|_{\mathrm{F}}^2+\beta_1 \sum_{i=1}^N \log(u_i+\epsilon_0)   \\
     & \mbox{s.t.} \;\;\;\norm{\vec{x}_i} \leq u_i ,~ \forall i \in \mathcal{N}.
 \end{array}
   \label{eq::log-sum}
\end{equation}
The log-sum penalty resembles most closely  the $\ell_0$-norm penalty when $\epsilon_0 \!\rightarrow 0$. However, a practical choice is to set $\epsilon_0$  to be slightly less than the expected amplitude of the non-zero rows in $\vec{X}\tran$ \cite{candes2008enhancing}. 

The optimization problem in \eqref{eq::log-sum} is a sum of a convex and a concave functions, thus, it is not convex in general. Therefore, we rely on majorization-minimization (MM) approach and we  approximate the concave penalty by its first-order Taylor expansion. Subsequently, we solve  \eqref{eq::log-sum} as an iterative reweighted problem given as
\vspace{-.05cm}
\begin{equation}\label{eq::MM_x}\footnotesize
\vec{X}^{(l+1)}=\displaystyle\min_{\vec{X}}\displaystyle  \frac{1}{2} \big\| \vec{\Phi}\vec{X}\tran - \vec{Y} \big\|_{\mathrm{F}}^2+\beta_{1} \sum_{i=1}^{N} g_{i}^{(l)}\|\vec{x}_{i}\|_{2},
\end{equation}
where $(l)$ denotes the MM iteration and 
$g_{i}^{(l)} = (\epsilon_0+\| \vec{x}_{i}^{(l)}\|_{2})^{-1}$.

 The objective function in problem  \eqref{eq::MM_x} is convex and it can
be solved optimally utilizing standard convex optimization
techniques. However, as the mMTC system  may grow large, the standard techniques  may not be computationally efficient.  Thus,  we propose the use of  ADMM to solve the optimization  
problem in \eqref{eq::MM_x} at each MM iteration $(l)$. 


Specifically,  we introduce an auxiliary variable  ${\vec{Z}\in \mathbb{C}^{M\times N}}$ and the dual variable $\vec{\Lambda} \in \mathbb{C}^{M\times N}$, hence,   the  augmented Lagrangian associated with \eqref{eq::MM_x}  is given by \begin{equation}\footnotesize
\min_{{\vec{X},\vec{Z}}}
\beta_1 \displaystyle\sum_{i=1}^N g_i^{(l)} \norm{\vec{x}_i} +  \displaystyle\frac{1}{2}
 \| \vec{\Phi} \vec{Z}\tran -\vec{Y}\|_{\mathrm{F}}^2  +\displaystyle\frac{\rho}{2}\|\vec{X}-\vec{Z}+\displaystyle\frac{\vec{\Lambda}}{\rho}\|_{\mathrm{F}}^2-\displaystyle\frac{\fro{\vec{\Lambda}}}{2\rho},
 \label{eq::Lagrange_ADMM}\end{equation}
 where $\rho$ is a positive parameter. The ADMM solves the optimization problem through sequential updates of  $(\vec{Z},\vec{X},\vec{\Lambda})$ as follows \cite{boyd2011distributed}:
\begin{equation}\footnotesize
 \vec{Z}^{(k+1)}:=\min_{\vec{Z}}\frac{1}{2}\|
\vec{\Phi}\vec{Z}\tran -\vec{Y}\|_{\mathrm{F}}^2+ \frac{\rho}{2} \Vert   \vec{X}^{(k)} - \vec{Z} +\frac{1}{\rho}\vec{\Lambda}^{(k)} \|_{\mathrm{F}}^2
\label{eq::z(k+1)}
\end{equation}
\begin{equation}\footnotesize
 \vec{X}^{(k+1)}:=\min_{\vec{X}}  \sum_{i=1}^{N} \beta_1 g_i^{(l)} \Vert  \vec{x}_i\|_2   +\frac{\rho}{2} \|\vec{X}-\vec{Z}^{(k+1)}+ \frac{1}{\rho}\vec{\Lambda}^{(k)} \|_{\mathrm{F}}^2
\label{eq::x(k+1)}
\end{equation}
  \begin{equation}\footnotesize
\vec{\Lambda}^{(k+1)} := \vec{\Lambda}^{(k)}+\rho\big(\vec{X}^{(k+1)}  -\vec{Z}^{(k+1)} \big),
\label{eq::lambda(k+1)}
\end{equation}
where the superscript $(k)$ denotes the ADMM iteration index. 
The derivations of the ADMM steps \eqref{eq::z(k+1)} and \eqref{eq::x(k+1)} are detailed  below. 

The $\vec{Z}$-update step in \eqref{eq::z(k+1)} solves a convex optimization problem. Thus,  $\vec{Z}^{(k+1)}$ is obtained by setting the gradient of the objective function in \eqref{eq::z(k+1)} with respect to $\vec{Z}$ to zero, resulting in
\begin{equation}\footnotesize
    \vec{Z}^{(k+1)}=\big(\rho \vec{X}^{(k)} +\vec{\Lambda}^{(k)}+\vec{Y}\tran\vec{\Phi}^*\big) \big( \vec{\Phi}\tran \vec{\Phi}^*+\rho \vec{I}_N\big)^{-1},
\end{equation}
where $(\cdot)^*$ denotes the complex conjugate operator.
 Note that the inversion $\big( \vec{\Phi}^*\vec{\Phi}\tran+\rho \vec{I}_N\big)^{-1}$ can be computed once
and stored to expedite the $\vec{Z}$-update step. 

Next,  the $\vec{X}$-update in \eqref{eq::x(k+1)} can be decomposed into $N$ sub-problems as follows
\begin{equation}\footnotesize
  \vec{x}_i^{(k+1)}:=\min_{\vec{x}_i} \displaystyle\frac{\beta_1 g_i^{(l)}}{\rho}\|  \vec{x}_i\|_2   +\frac{1}{2} \|\vec{x}_i-\vec{c}_i^{(k)}\|_2^2, \;\;\; \forall i \in \mathcal{N},
  \label{eq::l1}
\end{equation}
where $\vec{c}_i^{(k)}=\vec{z}_i^{(k+1)}- \dfrac{1}{\rho}\vecgreek{\lambda}_i^{(k)}$ and $\vecgreek{\lambda}_i^{(k)}$ is the $i$th column of $\vec{\Lambda}^{(k)}$.  The problem in \eqref{eq::l1} admits the closed-form  solution given by\cite{goldstein2014field}
\begin{equation}\footnotesize
    \vec{x}_i^{(k+1)}= \frac{\max{\Big\{0,\|\vec{c}_i^{(k)} \|_2-\frac{\beta_1 g_i^{(l)}}{\rho}\Big\}}}{\|\vec{c}_i^{(k)} \|_2}\vec{c}_i^{(k)},\quad \forall i \in \mathcal{N}.
    \label{eq::prox}
\end{equation}

\subsection{Covariance Aided JUICE}
Although the sparsity of the  matrix $\vec{X}$ is utilized in \eqref{eq::log-sum}, the  information  embedded in channel covariance matrices available at the BS is neglected. On this account, we reformulate the  problem in \eqref{eq::log-sum} so that it exploits also the covariance information. The key idea is that the sample covariance matrix $\vec{x}_i\vec{x}_i\herm$  for each active UE $i\in \mathcal{S}$ carries similar information as the true scaled covariance matrix $\tilde{\vec{R}}_i= \rho_i\vec{R}_i$.

Based on the above arguments, we augment the optimization problem  \eqref{eq::log-sum} with a regularization  term that penalizes the deviation of the sample covariance matrix $\vec{x}_i \vec{x}_i\herm$ from the true scaled covariance matrix $\tilde{\vec{R}}_i$. Thus, the covariance aided  JUICE problem is expressed as follows
\begin{equation}\footnotesize\vspace{-.02cm}
\begin{array}{ll}
&\displaystyle\min_{\vec{X},\vec{u} }\displaystyle\frac{1}{2}\|\vec{\Phi} \vec{X}\tran - \vec{Y}\|_{\mathrm{F}}^2+\beta_1 \sum_{i=1}^N \log(u_i+\epsilon_0)  \\
     & +\beta_2 \displaystyle\sum_{i=1}^{N} \mathbb{I}(u_{i}) \| \vec{x}_i\vec{x}_i\herm-\tilde{\vec{R}}_i  \| _{\mathrm{F}}^2 \;\;\; \mbox{s.t.} \;\;\;\norm{\vec{x}_i} \leq u_i ,~ \forall i \in \mathcal{N},
   \end{array} \label{eq:MMV}
\end{equation}
where $\beta_2$ controls the penalty on the covariance deviation term and $\mathbb{I}(\cdot)$ is an indicator function given by
\begin{equation}
  \footnotesize
\mathbb{I}(u_{i})=
\begin{cases}
1,& u_i > 0\\
0,&u_i= 0.
\end{cases}\label{eq::I(X)}
\end{equation}

Note that $\mathbb{I}(u_{i})$ ensures that only the estimated active UEs are penalized with the covariance regularization term. The indicator function is hard to handle due to its   \emph{combinatorial} nature. Therefore, we relax $\eqref{eq::I(X)}$ with a function $f(\cdot)$ that approximates the sign step function for positive values $v$, i.e., we define 
\begin{equation}\footnotesize
\label{eq:compander}
f(v;\kappa)=\frac{\log\big(1+\kappa v\big)}{\log(1+\kappa)},
\end{equation}
where $\kappa$ is a positive parameter to adjust the steepness of the function for small input values\cite{sriperumbudur2011majorization}. Subsequently,  \eqref{eq:MMV} is relaxed as 
\begin{equation}\footnotesize
\begin{array}{ll}
&\displaystyle\min_{\vec{X},\vec{u} }\displaystyle\frac{1}{2}\|\vec{\Phi} \vec{X}\tran - \vec{Y}\|_{\mathrm{F}}^2+\beta_1 \sum_{i=1}^N \log(u_i+\epsilon_0)  \\
     & +\beta_2 \displaystyle\sum_{i=1}^{N} f(u_{i};\kappa) \| \vec{x}_i\vec{x}_i\herm-\tilde{\vec{R}}_i  \| _{\mathrm{F}}^2 \;\;\; \mbox{s.t.} \;\;\;\norm{\vec{x}_i} \leq u_i ,~ \forall i \in \mathcal{N}.
 \label{eq:MMV_approx}
   \end{array}
\end{equation}

Since both the log-sum penalty and $f(u_i;\kappa)$ are concave functions, we rely on the MM approach and we approximate  the problem in  \eqref{eq:MMV_approx}  by its first-order Taylor expansion at $\vec{u}^{(l)}$. Subsequently, with the use of some simple manipulations, we can solve \eqref{eq:MMV_approx} as the following iterative reweighted problem given at $l$th MM iteration by
\begin{equation}\footnotesize
\label{eq:optProb}
\begin{array}{ll}
 \vec{X}^{(l+1)}= &\displaystyle\min_{\vec{X}}\displaystyle \frac{1}{2} \big\| \vec{\Phi}\vec{X}\tran- \vec{Y} \big\|_{\mathrm{F}}^2+\beta_{1}\sum_{i=1}^{N} g_{i}^{(l)}\|\vec{x}_{i}\|_{2}  \\ & +\beta_{2}\displaystyle \sum_{i=1}^{N}q_{i}^{(l)}\|\vec{x}_{i}\|_{2} \big\|\vec{x}_{i}\vec{x}_{i}\herm - \tilde{\vec{R}}_{i} \big\|_{\mathrm{F}}^2 \\
  \end{array}
\end{equation}
with $q_{i}^{(l)} = \displaystyle\frac{\kappa}{\log(1+\kappa)}\frac{1}{1+\kappa{\|\vec{x}_{i}^{(l)}\|_{2}}},~\forall i \in \mathcal{N}$.


The objective function in \eqref{eq:optProb} is \emph{non-convex} due  to the  covariance deviation penalty term. Therefore, in order to overcome the non-convexity, we introduce the splitting variables $\vec{Z},\vec{V} \in \mathbb{C}^{M\times N}$ and we rewrite the objective function in  \eqref{eq:optProb} as 
\begin{equation}\footnotesize
\label{eq:optProb_cvx}
\hspace{-.75cm}
\begin{array}{ll}
&\displaystyle\min_{\vec{X}}\displaystyle \frac{1}{2} \big\| \vec{\Phi}\vec{Z}\tran - \vec{Y} \big\|_{\mathrm{F}}^2 +\beta_{1}\sum_{i=1}^{N} g_{i}^{(l)}\|\vec{x}_{i}\|_{2} \\&+\beta_{2}\displaystyle \sum_{i=1}^{N}q_{i}^{(l)}\|\vec{x}_{i}\|_{2} \big\|\vec{z}_{i}\vec{v}_{i}\herm - \tilde{\vec{R}}_{i} \big\|_{\mathrm{F}}^2 \\ &
      \;\;\; \mbox{s.t.} \;\;\;  \quad\vec{x}_i=\vec{z}_i,\;\; \vec{x}_i=\vec{v}_i,~\forall i \in \mathcal{N}.
  \end{array}
\end{equation}


The optimization problem in \eqref{eq:optProb_cvx} is \emph{block multi-convex}, thus, we utilize ADMM to solve it efficiently. Accordingly,
the augmented Lagrangian  associated with \eqref{eq:optProb_cvx}  is given by
\begin{equation}\label{eq::lagrange_cov}
\footnotesize
\begin{array}{ll}
&\displaystyle\min_{\vec{Z},\vec{V},\vec{X}}  \frac{1}{2}
 \| \vec{\Phi} \vec{Z}\tran -\vec{Y}\|_{\mathrm{F}}^2 + 
 \beta_2  \displaystyle\sum_{i=1}^{N} q_{i}^{(l)}\Vert \vec{x}_i \|_{2}\Vert \vec{z}_i\vec{v}_i\herm - \tilde{\vec{R}}_i  \|_{\mathrm{F}}^2\\&+\displaystyle\sum_{i=1}^{N}\beta_{1} g_{i}^{(l)}\|\vec{x}_{i}\|_{2}  +\displaystyle\frac{\rho}{2}  \|\vec{X}- \vec{Z}  +\displaystyle\frac{\vecgreek{\Lambda}_{\mathrm{z}}}{\rho}\|_{\mathrm{F}}^2  +\displaystyle\frac{\rho}{2} \|\vec{X}- \vec{V}  +\displaystyle\frac{\vecgreek{\Lambda}_{\mathrm{v}}}{\rho}\|_{\mathrm{F}}^2\\&\displaystyle-\frac{ \fro{\vecgreek{\Lambda}_{\mathrm{v}}}}{2\rho}-\frac{\fro{\vecgreek{\Lambda}_{\mathrm{z}}}}{2\rho},
  \end{array}
\end{equation}
where $\vecgreek{\Lambda}_{\mathrm{z}}=[\vecgreek{\lambda}_{\mathrm{z}_1},\ldots,\vecgreek{\lambda}_{\mathrm{z}_N}]$ and $\vecgreek{\Lambda}_{\mathrm{v}}=[\vecgreek{\lambda}_{\mathrm{v}_1},\ldots,\vecgreek{\lambda}_{\mathrm{v}_N}]$ are the ADMM dual variables. Similarly to (\ref{eq::z(k+1)})--(\ref{eq::lambda(k+1)}), ADMM updates sequentially the primal variables $\vec{Z}$, $\vec{V}$, and $\vec{X}$ then the dual variables $\vecgreek{\Lambda}_{\mathrm{z}}$ and $\vecgreek{\Lambda}_{\mathrm{v}}$.

First, the $\vec{Z}$-subproblem, i.e., minimizing \eqref{eq::lagrange_cov} with respect to $\vec{Z}$, is given by
\begin{equation}\footnotesize
\begin{array}{ll}
\hspace{-.2cm}
\vec{Z}^{(k+1)}:=\!\!\!\!\!\!&\displaystyle\min_{\vec{Z}} \displaystyle\frac{1}{2}
 \|\vec{\Phi} \vec{Z}\tran -\vec{Y} \| _{\mathrm{F}}^2 + \displaystyle\frac{\rho}{2} \|\vec{X}^{(k)}- \vec{Z}  +\displaystyle\frac{\vecgreek{\Lambda}_{\mathrm{z}}^{(k)}}{\rho}\|_{\mathrm{F}}^2\\
     & + \beta_2 q_{i}^{(l)}  \displaystyle\sum_{i=1}^{N} \norm{\vec{x}_i^{(k)}} \| \vec{z}_i\vec{v}_i^{(k)\herm} - \tilde{\vec{R}}_i \|_{\mathrm{F}}^2. 
\end{array}
\label{eq::Z(k+1)}
\end{equation}
The objective function in  \eqref{eq::Z(k+1)}  is convex and the solution is obtained by setting the gradient with respect to $\vec{Z}$ to zero, resulting in
\begin{equation}\footnotesize
       \vec{Z}^{(k+1)}=\big(\vec{Y}\tran\vec{\Phi}^*+\vec{B}^{(k)}\big) \big(\vec{\Phi}\tran \vec{\Phi}^*+\vec{D}^{(k)}\big)^{-1}.
       \label{eq::z_k+1}
\end{equation}
where $\vec{b}_i^{(k)}=2 \beta_2q_{i}^{(l)} \norm{\vec{x}_i^{(k)}}\tilde{\vec{R}}_i\vec{v}_i^{(k)}+\rho \vec{x}_i^{(k)}+\vecgreek{\lambda}_{\mathrm{z}_i}^{(k)}$ is the $i$th column of matrix $\vec{B}^{(k)}$ and the matrix $\vec{D}^{(k)}$ is a diagonal matrix with entries $d_i^{(k)}=2 \beta_2 q_{i}^{(l)}\norm{\vec{x}_i^{(k)}}\norm{\vec{v}_i^{(k)}}^2+\rho$.

Second, the $\vec{V}$-update solves the  minimization problem given by
\begin{equation}\footnotesize
\displaystyle\min_{\vec{V}}  \beta_2  \displaystyle\sum_{i=1}^{N}q_{i}^{(l)} \norm{\vec{x}_i^{(k)}} \|\vec{z}_i^{(k+1)}\vec{v}_i\herm - \tilde{\vec{R}}_i\| _{\mathrm{F}}^2+\displaystyle\frac{\rho}{2} \| \vec{X}^{(k)}- \vec{V}  +\displaystyle\frac{\vecgreek{\Lambda}_{\mathrm{v}}^{(k)}}{\rho}\|_{\mathrm{F}}^2. 
\label{eq::v(k+1)}
\end{equation}
The optimization problem in  \eqref{eq::v(k+1)} can be decoupled into $N$ convex sub-problems, with a unique solution given by:
\begin{equation}\footnotesize
       \vec{v}_i^{(k+1)}=\frac{2\beta_2 q_i^{(l)}  \norm{\vec{x}^{(k)}_i} \tilde{\vec{R}}_i\vec{z}^{(k+1)}_i+\rho \vec{x}_i^{(k)}+ \vecgreek{\lambda}_{\mathrm{v}_i}^{(k)}}{2\beta_2 q_i^{(l)}  \| \vec{x}^{(k)}_i \|_2\norm{\vec{z}^{(k+1)}_i}^2 +\rho},\,\, \forall i \in \mathcal{N}.
       \label{eq::v_k+1}
\end{equation}

Next, with some manipulations, the $\vec{X}$-update solves the following convex optimization problem
\begin{equation}\footnotesize
\begin{split}
    \vec{X}^{(k+1)}:=&\min _{\vec{X}} \sum_{i=1}^{N} \alpha_i^{(k)} \norm{\vec{x}_i} +\rho  \|   \vec{X}-  \vec{S}^{(k)}\| _{\mathrm{F}}^2,  
  \end{split}
  \label{eq::min_X}
\end{equation}
where $\vec{S}^{(k)}=\dfrac{1}{2}\big( \vec{Z}^{(k+1)}+\vec{V}^{(k+1)}-\displaystyle\frac{\vecgreek{\Lambda}_{\mathrm{z}}^{(k)}+\vecgreek{\Lambda}_{\mathrm{v}}^{(k)}}{\rho}\big)$ and $\alpha_i^{(k)}= \beta_1 g_i^{(l)}+\beta_2 q_i^{(l)}\| \vec{z}_i^{(k+1)}\vec{v}_i^{(k+1)\herm}- \tilde{\vec{R}}_i\|_{\mathrm{F}}^2$. The optimal solution to  \eqref{eq::min_X}  has a closed-form expression given by
\begin{equation}\footnotesize
\vec{x}_i^{(k+1)}= \frac{\max{\big\{0,\norm{\vec{s}_i^{(k)}}-\frac{\alpha_i^{(k)}}{2\rho}\big\}}}{\norm{\vec{s}_i^{(k)}}}\vec{s}_i^{(k)},\quad \forall i \in \mathcal{N}.
    \label{eq::prox2}
\end{equation}


 
 The details of the proposed covariance aided JUICE, termed  as cov-ADMM, are summarized in Algorithm 1.  Note that if the second-order channel statistics are not available, we set $\beta_2=0$, hence, Algorithm 1 presents  the proposed iterative reweighted ADMM (IRW-ADMM) in Section \ref{IRW-ADMM}. Moreover, if $\beta_2=0$ and $q_i^{(l)}=g_i^{(l)}=1,$ for $i\in \mathcal{N}, ~l=1,2,\ldots$,  Algorithm 1 presents the ADMM  solution, which we call ADMM, for the  problem in \eqref{eq::l_1}.

\begin{algorithm}[h]
\DontPrintSemicolon
   \KwInput{$\{\tilde{\vec{R}}_i\}_{i=1}^N,\beta_1,\beta_2,\rho,\epsilon_0,\epsilon,\kappa$}
  \KwOutput{$\hat{\vec{X}}$}
 \Kwinitialize{$\vec{X}^{(0)},\vec{V}^{(0)},\vec{Z}^{(0)},\vecgreek{\Lambda}_{\mathrm{v}}^{(0)},\vecgreek{\Lambda}_{\mathrm{z}}^{(0)}.$}
   \While{$l<l_{\mathrm{max}}$  }
   {
   \While{$k<k_{\mathrm{max}}$ $\mathrm{or}$ $\| \vec{X}^{(k)}-\vec{X}^{(k-1)} \|<\epsilon$}
   {
Update $\vec{Z}^{(k+1)}$ using equation \eqref{eq::z_k+1}\;
 Update $\vec{V}^{(k+1)}$ using equation \eqref{eq::v_k+1}\;
Update $\vec{X}^{(k+1)}$ using equation \eqref{eq::prox2}\;
 $\footnotesize\vecgreek{\Lambda}_{\mathrm{z}}^{(k+1)}=  \vecgreek{\Lambda}_{\mathrm{z}}^{(k)}+\rho\big(   \vec{X}^{(k+1)}-\vec{Z}^{(k+1)} \big)$\;
 $\footnotesize\vecgreek{\Lambda}_{\mathrm{v}}^{(k+1)}=  \vecgreek{\Lambda}_{\mathrm{v}}^{(k)}+\rho\big(   \vec{X}^{(k+1)}-\vec{V}^{(k+1)} \big)$\;
   $k\leftarrow{k+1}$\;

   }
  $ g_{i}^{(l)} = (\epsilon_0+\| \vec{x}_{i}^{(l)}\|_{2})^{-1}, i\in\mathcal{N}$ \;
$q_{i}^{(l)} = \displaystyle\frac{\kappa}{\log(1+\kappa)}\frac{1}{1+\kappa{\|\vec{x}_{i}^{(l)}\|_{2}}},~\forall i \in \mathcal{N}$\;
$l\leftarrow{l+1}$\;
}
\caption{Covariance aided JUICE}
\end{algorithm}

\begin{figure*}[h!]
    \centering
     \begin{subfigure}{0.33\textwidth}
    \includegraphics[scale=.39]{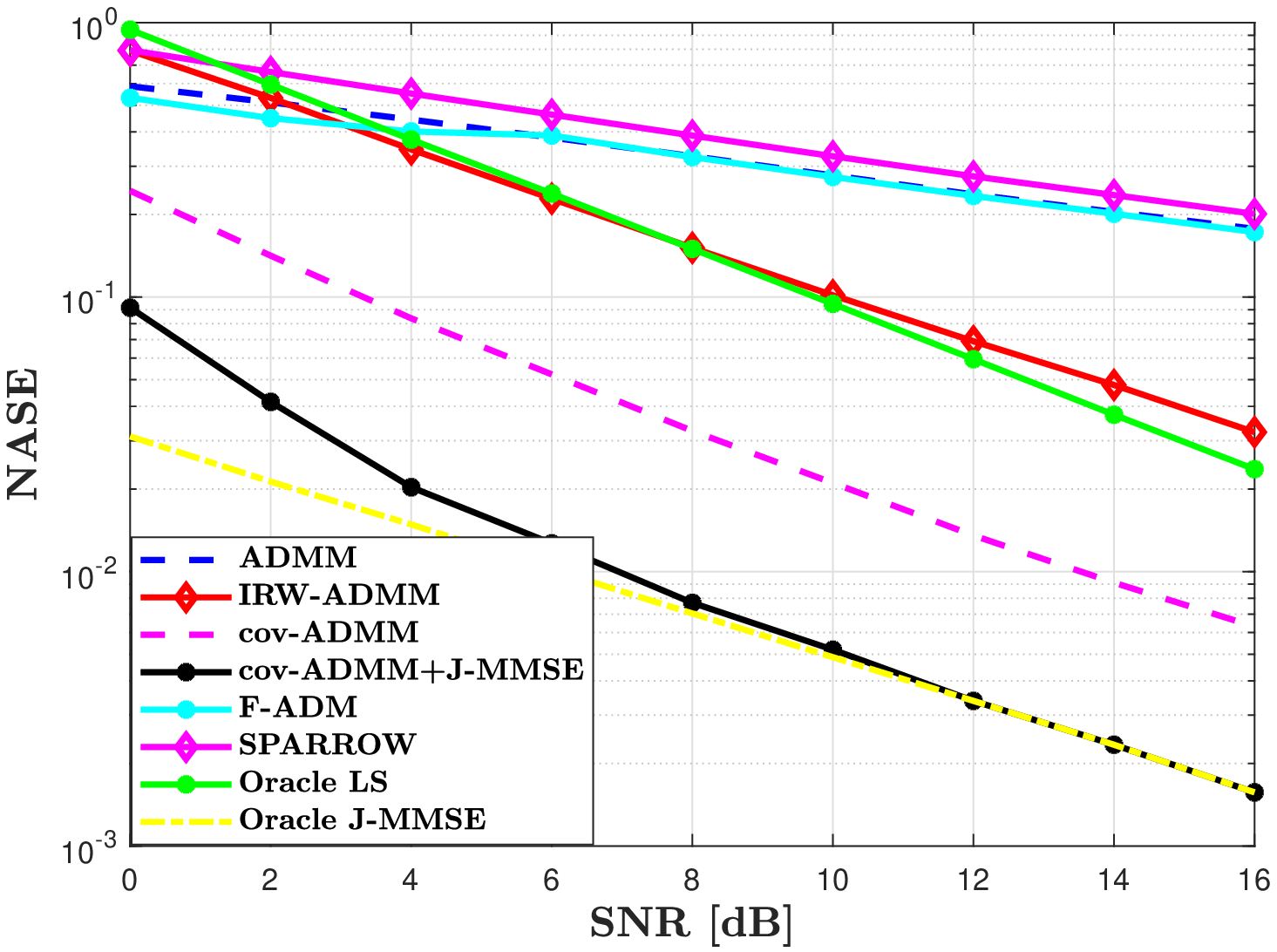}
    \caption{ NASE versus  SNR.}
    \label{fig:NASE}
\end{subfigure}
   \hfill      
   \begin{subfigure}{0.33\textwidth}
  \centering
     \includegraphics[scale=.39]{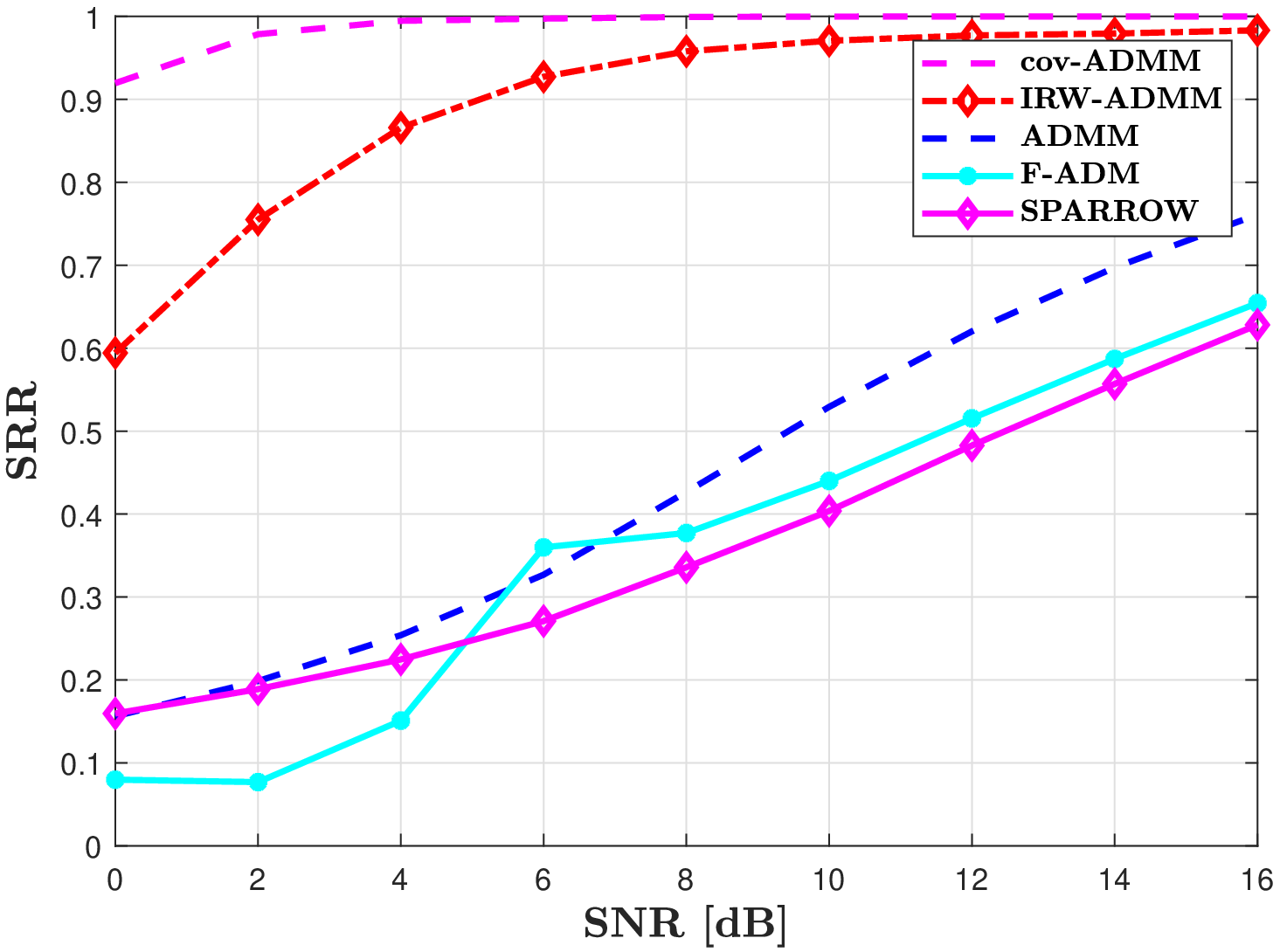}
     \caption{SRR rate versus SNR.}
     \label{fig:SR}
\end{subfigure}  
\hfill
    \begin{subfigure}{0.33\textwidth}
  \centering
     \includegraphics[scale=.39]{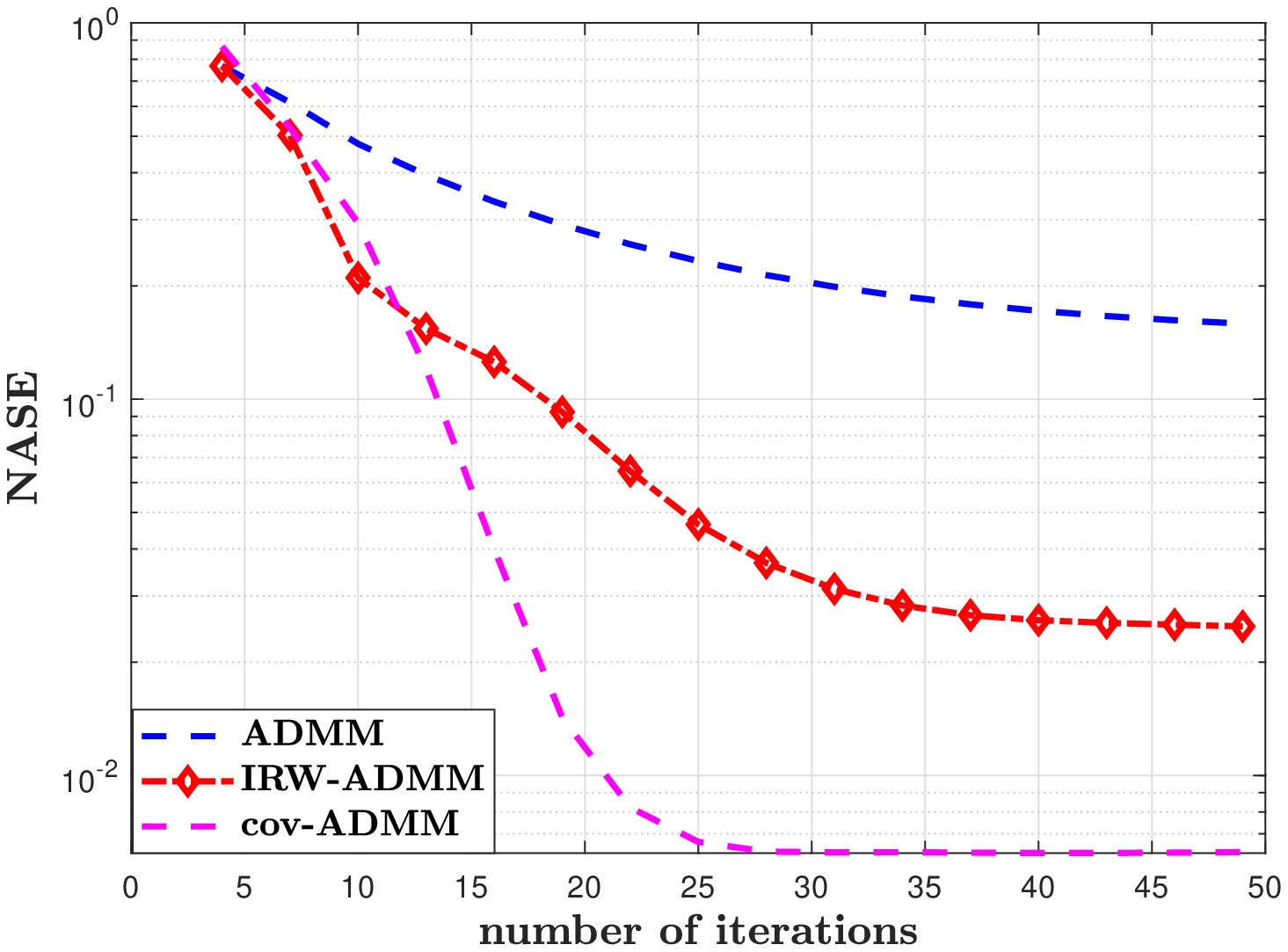}
     \caption{NASE versus the number of iterations.}
     \label{fig:iter_mmv}
\end{subfigure}\vspace{-2mm}
\caption{ Performance comparison for the different algorithms for $N=200$, $M=20$, $K=10$, and $\tau_{\mathrm{p}}=20$.}
\label{fig:results}
\vspace{-3mm}
\end{figure*}

\subsection{MMSE-Based Channel Estimation}
The  estimated effective channel matrix of the active UEs, $\hat{\Vec{X}}_\mathcal{S}$,  provided by the proposed approach can be used for coherent data detection. However, if the second-order channel  statistics are known to the BS,  a more accurate channel estimate can be obtained  by applying the MMSE estimator.

Let us define the $ \mathrm{vec}(\cdot)$ operation as the column-wise stacking of a matrix. We define   $\vec{y}=\mathrm{vec}(\vec{Y}\tran) \in \mathbb{C}^{\tau_{\mathrm{p}} M}$,  $\vec{w}=\mathrm{vec}(\vec{W}\tran) \in \mathbb{C}^{ \tau_{\mathrm{p}}M} $ and $\vec{x}=\mathrm{vec}(\vec{X}_\mathcal{S}) \in \mathbb{C}^{KM}$. Accordingly, we rewrite \eqref{eq::Y_matrix}  as\footnote{We assume perfect UEs identification, as the main goal is to show channel estimation quality improvement gained by using the  MMSE estimator.}: 
 \begin{equation}\footnotesize
    \vec{y}=\vec{\Theta}  \vec{x}+  \vec{w},
    \label{Y_vec}
 \end{equation}
 where $\vec{\Theta}=\vecgreek{\Phi}_\mathcal{S}\otimes\vec{I}_{M} \in \mathbb{C}^{M\tau_{\mathrm{p}} \times KM}$, and the operator $\otimes$  denotes the Kronecker product.  The vectorization in \eqref{Y_vec}  transforms the matrix estimation into a classical form of vector estimation which enables the
use of the linear MMSE estimator  given by \cite[Eq.~(12.26)]{kay1993fundamentals}
\begin{equation}\footnotesize
     \vec{x}^{\mathrm{J-MMSE}}=\mathrm{vec}(\vec{X}^{\mathrm{J-MMSE}})=\bar{\vec{x}}+\vec{R}_{\mathrm{diag}}\vec{\Theta}\herm \vec{Q}\big(\vec{y}-\vec{\Theta}\bar{\vec{x}}\big),\label{eq::mmse}
 \end{equation}
where $\vec{Q}=( \vec{\Theta}\vec{R}_{\mathrm{diag}}\vec{\Theta}\herm+\sigma^2 \vec{I}_{\tau_{\mathrm{p}}M} )^{-1}$, $\bar{\vec{x}}$ denotes the mean of $\vec{x}$, and $\vec{R}_{\mathrm{diag}}$ denotes the covariance matrix of  $\vec{x}$ given as a block diagonal matrix with the main-diagonal blocks are given by the scaled covariance matrices $\tilde{\vec{R}}_i$ corresponding to the active UEs $i \in \mathcal{S}$.

\vspace{-.2cm}
\section{Numerical Results}
\label{Result}

We consider a single-cell network that consists of one BS equipped with $M=20$ antennas serving a total of  $N=200$ uniformly distributed  UEs, out of which only $K=10$  are active at each $T_\mathrm{c}$. For the channel model in \eqref{eq::ch}, we consider $P_i=200, ~ \forall i\in \mathcal{N}$,  and the  mean AoA for each UE is uniformly distributed  over $ [\frac{\pi}{3},\frac{2\pi}{3}]$. Each user  $i \in \mathcal{N}$ is assigned  a unique normalized quadratic phase-shift keying  sequence $\vecgreek{\phi}_i$, with $\tau_{\mathrm{p}}=20$, generated from an i.i.d. complex Bernoulli distribution.

Channel estimation is quantified in terms of normalized mean square error (NMSE) defined as  $\frac{\mathbb{E}\left [\| \vec{X}-\hat{\vec{X}}_\mathcal{S}\|_{\mathrm{F}}^2 \right ]}{\mathbb{E}\left[\| \vec{X}\|_{\mathrm{F}}^2 \right]},$
where the expectation is computed via Monte-Carlo averaging over all sources of randomness. Thus, the NMSE  is presented as the \textit{normalized average square error} (NASE). User activity detection is quantified in terms of  support recovery  rate (SRR) defined as $\frac{\vert \mathcal{S} \cap \hat{\mathcal{S}}\vert}{\vert \mathcal{S} - \hat{\mathcal{S}}\vert+K}, $
where $\hat{\mathcal{S}}$ denotes the detected support. 

We compare the performance of cov-ADMM, IRW-ADMM, and ADMM to two  algorithms that solve the problem in \eqref{eq::l_1}, namely, fast alternating direction methods (F-ADM)  \cite{lu2011fast} and  SPARROW \cite{steffens2018compact}. In addition, we use  genie-aided least square (LS) and genie-aided MMSE estimators that are provided ``oracle'' knowledge on the true set of active UEs to establish an optimal performance benchmark.



Fig. \ref{fig:results}(a) presents the channel estimation performance in terms of  NASE   against SNR. First, in the case when the second-order statistics of the channels are not available at the BS, the proposed IRW-ADMM  provides a significant improvement to the channel estimation quality compared to  ADMM, ADM, and SPARROW. Furthermore, IRW-ADMM achieves a similar   performance compared to the oracle LS estimator. This result points out clearly the remarkable gain obtained by  the iterative reweighted $\ell_{2,1}$-norm minimization approach. Second, if the BS is provided with the second-order statistics of the channels,  the proposed cov-ADMM improves considerably the channel estimation. In fact, it provides the same performance as IRW-ADMM while  using $10$ dB lower SNR.  Moreover, using the cov-ADMM with an MMSE estimator renders the same performance as the oracle MMSE estimator starting at SNR $=10$~dB.

Fig. \ref{fig:results}(b)  shows the user identification accuracy in terms of  SRR rate against SNR. The results show that cov-ADMM   indisputably provides the highest SRR rate amongst all the considered algorithms. In fact, cov-ADMM   identifies the set  of true active users perfectly for SNR $\geq 10 $~dB. In addition, the IRW-ADMM provides a significant improvement compared to ADMM, F-ADM, and SPARROW and it achieves an SRR rate $>0.95$ around SNR $=8$~dB. 

Fig. \ref{fig:results}(c)  shows the typical convergence  behavior of the proposed algorithms at  SNR $=16$~dB.  The results reveal that IRW-ADMM requires approximately $50$ iterations to convergence. We note that the early
iterations may find inaccurate signal estimates, hence, the lower performance when the number of iterations is less than 10. Furthermore,  the proposed cov-ADMM convergence to its optimal solution in about 25 iterations. The results presented in Fig. \ref{fig:results}  highlight clearly  the significant gains obtained by exploiting available  prior channel covariance information at the BS, as it yields the best performance in terms of  channel estimation, user detection, and  convergence rate.




 \vspace{-.2cm}

\section{Conclusion}

The paper investigated joint support and  signal recovery from an  MMV  model for the use case of user identification and  channel estimation in MIMO-based grant-free mMTC. The paper proposed the formulation of JUICE  based on an iterative reweighted $\ell_{2,1}$-norm minimization problem that exploits the second-order channel statistics when they are available to the BS. An  ADMM-based algorithm was derived to provide a computationally efficient solution. The numerical results show significant improvement in UEs activity detection accuracy, channel estimation quality, and   convergence rate.

\label{sec:refs}

\bibliographystyle{IEEEbib}
\bibliography{main}
\end{document}